\documentclass[preprint]{aastex}

\newcommand{\et}{et al.\ }
\newcommand{\xte}{{\it RXTE}}
\newcommand{\xmm}{{\it XMM-Newton}}

\newcommand{\chandra}{{\it Chandra}}
\newcommand{\ginga}{{\it Ginga}}

\newcommand{\asca}{{\it ASCA}}
\newcommand{\Msun}{\hbox{$\rm\thinspace M_{\odot}$}}
\newcommand{\mcg}{MCG{$-6$}-30-15}
\newcommand{\degg}{\hbox{$^\circ$}}
\newcommand{\ls}
{\mathrel{\hbox{\rlap{\hbox{\lower4pt\hbox{$\sim$}}}\hbox{$<$}}}} 
\newcommand{\gs}
{\mathrel{\hbox{\rlap{\hbox{\lower4pt\hbox{$\sim$}}}\hbox{$>$}}}} 
\def\Msun{\hbox{$\rm ~M_{\odot}$}}

\begin{document}
\title{Evidence for  Rapid Iron K$_{\alpha}$ Line Flux Variability in MCG--6-30-15}
\shorttitle{Rapid Iron Line Variability in MCG--6-30-15}
\shortauthors{Vaughan \& Edelson}
\author{Simon Vaughan}
\affil{X-Ray Astronomy Group; University of Leicester; Leicester, LE1 7RH; United Kingdom}
\authoremail{sav@star.le.ac.uk}
\and
\author{Rick Edelson}
\affil{Astronomy Department; University of California; Los Angeles, CA 90095-1562; USA}
\authoremail{rae@astro.ucla.edu}

\begin{abstract}

This  paper employs  direct spectral
fitting  of  individual  orbital  data in order to measure  rapid  X-ray iron
K$_{\alpha}$ line and continuum spectral slope variations in Seyfert 1
galaxies with unprecedented  temporal resolution.  Application of this
technique to  a long \xte\  observation of \mcg\ indicates  that the
line flux does vary on short ($\ls$1~d) timescales, but that these variations
are {\it not} correlated with  changes in the continuum flux or slope.
These rapid  variations indicate that  the line does  indeed originate
close  to the  black hole,  confirming predictions  based on  its very
broad  profile.   However,  the  lack  of  a  correlation  with  the
continuum presents problems for models in which the line
variations are driven by  those in the continuum, modified  only by light-travel
time effects.  Instead, it may  be that the line responds according to
a physical  process with  a different time  scale, such  as ionization
instabilities  in the  disk,  or  perhaps that  the  geometry and physical picture
is  more complex than implied by the simplest disk--corona models.

These data  also indicate that  the slope of the  underlying power-law
continuum ($\Gamma$) shows strong variability and is tightly correlated with the
continuum flux in the sense  that the spectrum steepens as the source brightens.
All  of these results  have been  checked with  extensive simulations,
which also indicated that  a spurious correlation between $\Gamma$ and
Compton reflection fraction ($R$)  will result if these quantities are
measured from the same spectra.  This casts serious doubts on previous
claims of such a $ \Gamma - R $ correlation.
\end{abstract}


\section{ Introduction }
\label{intro}

X-ray  observations probe  the  innermost regions  of Active  Galactic
Nuclei (AGN) and potentially offer a direct probe the immediate
environment of the central,  supermassive black holes thought to power
AGN  (e.g., Mushotzky, Done  \& Pounds  1993).  The  currently favored
model  for  the  central  regions  comprises a  hot  ($\sim  10^{8}  -
10^{9}$~K) optically-thin corona lying  above and below the surface of
an optically thick, geometrically thin accretion disk (e.g., Haardt \&
Maraschi 1993).  Ultraviolet photons from the disk are inverse-Compton
scattered to X-ray energies  (`up-scattered') as they pass through the
corona.  The  X-ray source, in  turn, illuminates the  accretion disk,
with  photons  either absorbed  and  thermalized  within  the disk  or
`reflected,'  producing a  characteristic reflection  hump  and strong
iron  K$_{\alpha}$ fluorescence  (Lightman  \& White  1988; George  \&
Fabian 1991).

\ginga\ observations showed  these features to be common  to Seyfert 1
nuclei and  provided strong support  for this model (Pounds  \et 1990;
Nandra  \& Pounds  1994).  However,  the  \asca\ mission 
provided  the   first  spectrally  resolved   iron  K$_{\alpha}$  line
profiles, showing that  many
Seyfert 1  galaxies to  have a broad,  redshifted iron line profiles  (Tanaka \et
1995; Nandra  \et 1997b), best  explained by a combination  of Doppler
and  gravitational redshifts from  the inner  regions of  an accretion
disk around  a black hole  (Fabian \et 1995).  These  provide
some of  the best  evidence in  support of the  black hole  model (see
reviews by Fabian \et 2000 and Nandra 2000).

Such a compact region {\it  should} respond very rapidly to changes in
the  illuminating flux  (Stella 1990;  Matt \&  Perola  1992).  Simple
models of X-ray illuminated accretion disks, in which the reprocessing
time  scale  is  very  short,  predict  the  line  should  respond  to
variations in the  continuum with a delay of  order the light-crossing
time ($\sim 1000 M_{7}$~s at  20$r_{g}$ from a black
hole of  mass $10^{7} M_{7}$\Msun, where $r_{g} = GM/c^{2}$).  The  response of the  line to the
continuum could then be used  to deduce the geometry and absolute size
scales of the inner accretion disk.  This could in principle provide a
more accurate method for estimating the mass of the central black hole
and,  when  combined with accurate  line  profiles,  even  reveal its  spin
(Reynolds \et 1999; Reynolds 2000).

In the  case of \mcg, the  profile is so  broad that most of  the line
flux  appears to  originate within  $\sim 20r_{g}$  of the  black hole
(Tanaka  \et 1995;  Iwasawa  \et 1996;  Guainazzi  \et 1999).   Recent
\chandra\ observations do not show any strong, narrow component to the
iron line (Lee \et 2000b), supporting the  view that, in
this object  at least, most of  the line flux originates  close to the
black hole.   Attempts to measure  the expected continuum  flux driven
variability of  the iron line (Iwasawa  \et 1996, 1999;  Lee \et 2000a)
have led to  suggestions that the line may  be variable, but also provide 
indications that  the simple picture  above may not fully  explain the
behavior.

The situation  in other Seyfert  1s is even less clear.   For instance,
Nandra \et  (1999) found  that in NGC~3516,  another Seyfert~1  with a
broad iron line,  the core of the line  profile (6.0--6.4~keV) appeared to track
the  continuum  but  the  broad   wings  of  the  line  varied  in  an
uncorrelated fashion.   In a long (30~d)  observation, NGC~7469 showed
marginal evidence for iron line variability correlated with X-ray flux
(Nandra \et 2000).  Complex variability  has also been reported in the
iron lines of NGC~4051 (Wang \et 1999) and NGC~7314 (Yaqoob \et 1996).
Weaver,  Gelbord  \&  Yaqoob   (2000) suggest  that  iron  line
variability is a common property in Seyfert 1s based on an analysis of
archival \asca\ data.
On the  other  hand,  Chiang  \et  (2000)  found  no  evidence  for  line
variability  in a  simultaneous \xte/\asca\  observation  of NGC~5548.
However, recent \chandra\ observations  of NGC~5548 revealed a strong,
narrow  component to  the  iron line,  indicating  that a  significant
fraction of the  line flux originates in distant  material (Yaqoob \et
2000).  A  constant, narrow  line is indeed  predicted  by AGN
unification schemes  (Krolik, 
Madau \& \.{Z}ycki 1994; Ghisellini, Haardt \& Matt 1994).

In order to determine if  Seyfert~1 iron lines do indeed vary rapidly,
we have  designed a new  analytical technique specifically  to measure
rapid changes in the X-ray  spectrum by direct spectral fitting to the
individual orbital data.  This has then been applied to \mcg, an ideal
subject  for this  approach because  it is  a bright,  highly variable
Seyfert 1 with an unambiguously broad iron line and a very long (8~d),
quasi-continuous \xte\ observation.

The rest  of the paper is  organized as follows.  In  the next section
the  observations and  data reduction  procedures are  described.  The
spectral analysis is discussed in \S~\ref{analysis} and the results of
this   analysis   are    tested   using   extensive   simulations   in
\S~\ref{simulations}.   The   implications  of  the   results  of  are
discussed in  \S~\ref{discussion}, followed by a brief  summary of the
main results in \S~\ref{conclusions}.

\section{ Observations and Data Reduction} 
\label{data}

\mcg\ was  observed by \xte\  during the period  1997 August 4  -- 12.
Further  details of  the observation and  an analysis  of the  
time-averaged spectral properties are given  by Lee \et (1999).  The \xte\
observations were taken simultaneously  with \asca\ (Iwasawa \et 1999;
Lee \et 2000a), but because the \asca\ count rate is a factor $\sim 10$
less than  that of \xte\  these data are  not as useful  for measuring
rapid spectral  variability.  The \asca\ data were  therefore not used
in the present analysis.

The \xte\ Proportional Counter Array (PCA) consists of five collimated
Proportional Counter  Units (PCUs), sensitive  to X-rays in  a nominal
2--60~keV   bandpass   and   with   a   total   collecting   area   of
$\sim6250$~cm$^2$.  PCU units  3 and 4 were turned  off during most of
the observation  due to performance  problems and here only  data from
three of the PCUs (0, 1  and 2) were used.  In the following analysis,
data  from  the top  (most  sensitive) layer  of  the  PCU array  were
extracted  using the  {\tt  REX} reduction  script  and {\tt  SAEXTRCT
v4.2b}.  Poor quality data were excluded on the basis of the following
acceptance criteria: the satellite has  been out of the South Atlantic
Anomaly   (SAA)  for   at  least   30  min;   Earth   elevation  angle
$\geq$~10\degg; offset from optical position of \mcg $\leq$~0.02\degg;
and {\tt ELECTRON0}$\leq$~0.1.  This  last criterion removes data with
high anti-coincidence  rate in the  propane layer of the  PCUs.  These
selection  criteria  left a  total  of  326~ks  of good  data.   The
background  was estimated using  the latest  version of  the `L7--240'
model   (Jahoda  \et   2000),  specifically   the  model   files  {\tt
pca\_bkgd\_faint240\_e3v19990909.mdl}             and             {\tt
pca\_bkgd\_faintl7\_e3v19990824.mdl} have been used\footnote{See {\tt
http://lheawww.gsfc.nasa.gov/$\sim$keith/dasmith/rossi2000/index.html}}. 

Light curves were  initially extracted from the {\tt STANDARD-2} data with 
16 s  time resolution.  The data were rebinned on orbital sampling
period (5760~s)  of \xte, with  the Earth-occultation gaps as  the bin
edges.  This  has the advantage  of improving the  signal-to-noise and
sampling the data on  the shortest available uninterrupted timescale.
A total of 123 orbits of  data were extracted.  Three of these contain
less than  1~ks of good  data and less than  $10^{4}$~source counts,
and were excluded on this basis.  The remaining 120 `good' orbits each
contain  $1.4 -  3.7$~ks  of good  exposure time and $1.2  - 6.1  \times
10^{4}$ source  counts.  Source and  background spectra
were extracted from each of  these 120 orbits.  Response matrices were
generated throughout  the length of the observation  using {\tt PCARSP
v2.43}, but  over such a short  duration the response  does not change
significantly.  In  the following spectral analysis  a response matrix
taken from the middle of the observation was used.

\section{ Spectral Analysis } 
\label{analysis}

The goal of this analysis  is to measure the variability properties of
the  iron  line  and  continuum  with the  highest  possible  temporal
resolution.   As discussed  above, the  natural sampling  rate  is the
satellite orbital period (96~min), as  that is the shortest timescale
on which no regular interruptions  in the light curve occur.  Once the
data were  individually extracted and  calibrated for each of  the 120
good  orbits, they  were  then  fitted using  the  {\tt XSPEC}  v11.01
(Arnaud  1996) software  package.   The key  to  keeping this  process
robust was to utilize the  minimum number of free parameters needed to
produce  an acceptable  fit (e.g.,  $\chi^2_\nu \approx  1 $)  in each
orbit.   The process  of determining  what  spectral model  to use  is
discussed in  the next two  subsections, and the errors  are estimated
and  final  data  are  presented  in the  following  two  subsections.
Simulated data were then used to test this approach under a variety of
conditions, as discussed in the following section.

\subsection{ Time-Averaged Spectral Fits } 
\label{fits}

\mcg\ is  known to  contain strong warm  absorber features  (Nandra \&
Pounds 1992; Reynolds \et 1995; George  \et 1998).  In order to reduce the
effect of this on the spectral analysis data below 3~keV were ignored
(see Figure~1 of Iwasawa \et 1996).
This  has  the additional  benefit  of  avoiding possible  calibration
problems with the  PCA below 3~keV.  The data  were further restricted
to energies less than 18~keV, above which the source count rate is too
low to obtain a good detection in each orbit.

\begin{figure}\vspace{22 cm}
\includegraphics{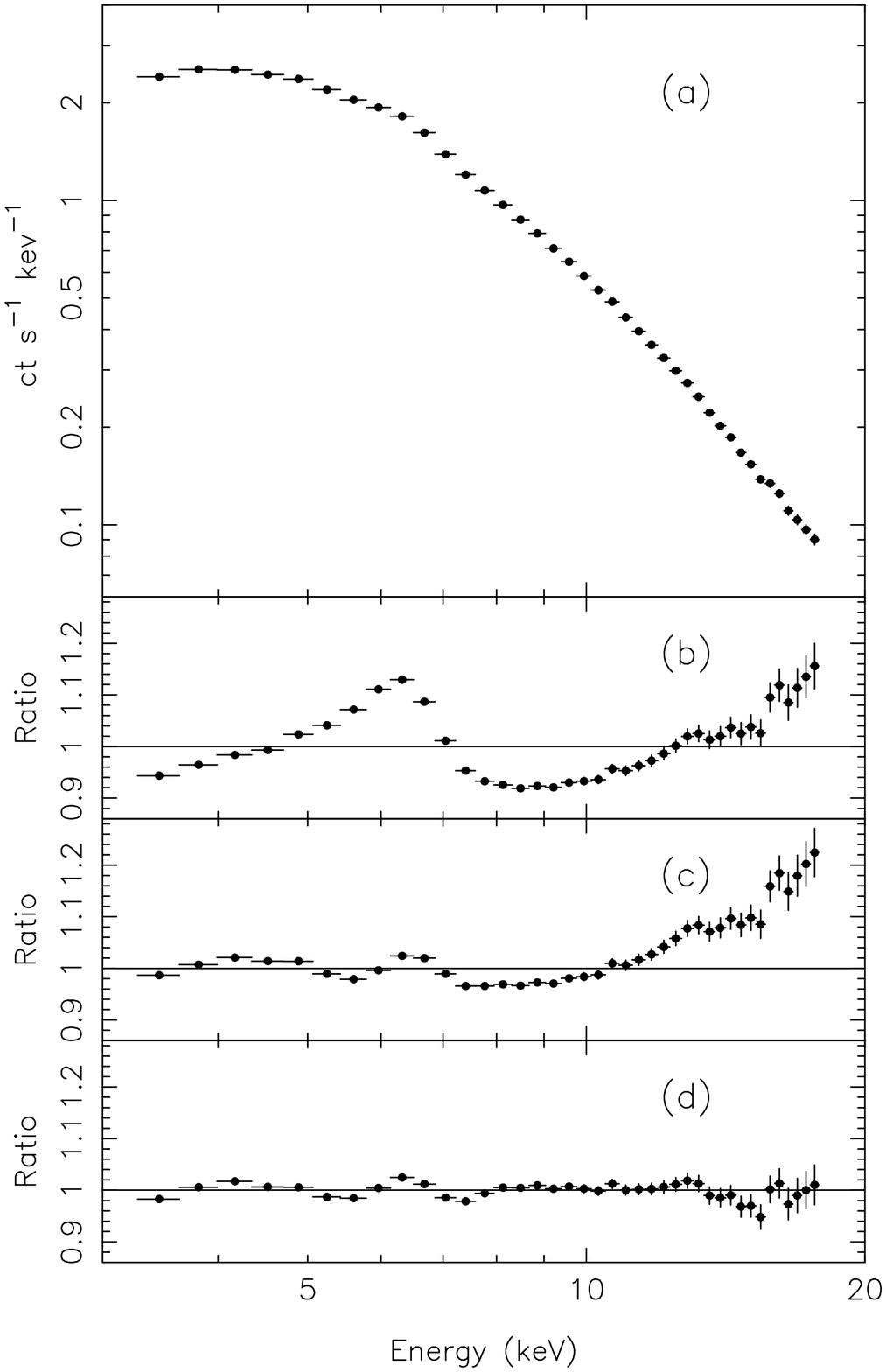}
\caption{ 
Time-averaged \xte\ PCA spectrum of \mcg. Panel $a$ shows the
count spectrum. Panels $b$, $c$ and $d$ show the data/model ratios
when compared to various models. Panel $b$ uses a simple power-law
model, panel $c$ uses a power-law plus broad Gaussian line and panel
$d$ uses a power-law, line and reflection component.}
\label{spectrum}
\end{figure}

The spectral  models were produced by fitting  the time-averaged \xte\
spectrum  (Figure~\ref{spectrum}).  In  each trial  model  the neutral
absorption column was fixed at the Galactic value, i.e., $N_{H} = 4.09
\times 10^{20}$~cm$^{-2}$, which Lee \et (1999) found to be consistent
with the \xte\ spectrum. (Such a small column has a negligible
effect on the spectrum above 3~keV.)  The time-averaged spectrum shows
clear    residuals    when   compared    to    a   simple    power-law
(Figure~\ref{spectrum}b).   The strongest  residual  feature peaks  at
$\sim 6$~keV and is due to the iron K$_{\alpha}$ line.  In addition to
this   the  spectrum   shows  a   significant  up-turn   above  10~keV
characteristic       of      a      Compton       reflection      hump
(Figure~\ref{spectrum}b,c).  Therefore the  first trial spectral model
(Figure~\ref{spectrum}d)  comprised  a  power-law,  a  broad  Gaussian
emission line  and a {\tt  pexriv} reflection component  (Magdziarz \&
Zdziarski 1995).  The inclination angle  of the reflector was fixed at
30\degg,  and the  $e$-folding energy  of the  incident  power-law was
fixed  at 100~keV,  consistent with  the previous  fits of  Tanaka \et
(1995), Guainazzi \et (1999) and Lee \et (1999).

This  model   provides  a   statistically  unacceptable  fit   to  the
time-averaged spectrum but, as  can be seen in Figure~\ref{spectrum}d,
the data/model residuals  are at the $\ls 3$~\%  level, which is about
the    limit     of    the    PCA     calibration\footnote{See    {\tt
http://lheawww.gsfc.nasa.gov/users/$\sim$keith/rossi2000/energy\_response.ps
}}.  Also,  the energy resolution of  the PCA ($\sim  1$~keV at 6~keV)
makes the PCA largely insensitive  to the detailed profile of the iron
line, and  therefore a  Gaussian is a  good approximation.   For these
reasons this  model was accepted  as a reasonable  parameterization of
the time-averaged spectrum.  The  fit parameters were: power-law slope
$\Gamma=2.06$;   line  energy   $E_{K\alpha}=5.73$~keV,  normalization
$F_{K\alpha}=2.17  \times 10^{-4}$  photons cm$^{-2}$  s$^{-1}$, width
$\sigma_{K\alpha}=0.87$~keV         and        equivalent        width
$EW_{K\alpha}=374$~eV; relative reflection strength $R=1.42$.

\subsection{ Time-Resolved Spectral Fits }

The above model  has been fitted to all 120  orbital spectra.  Even in
these short  observations, with only $\sim  2.5$~ks integration times,
the  line  was  always  well  detected.   As  with  the  time-averaged
spectrum, the  addition of a Gaussian line  significantly improved the
fits  to the  individual orbital  spectra  ($\gs 99.9$\%  level in  an
F-test) compared to a simple power-law model.  However, the reflection
component was  not nearly as well  detected in each  orbit (only $\sim
99$\% confidence).

The F-test was  used to decide what free parameters  to include in the
fit.  The spectral fits were  repeated with the parameters ($ \Gamma ,
F_{K\alpha} , R $) kept  fixed at their time-averaged values (as given
above).   The difference  in total  $\chi^2$ between  the fits  when a
given parameter was  fixed and when all parameters  were free was then
used  to assess  the significance  of allowing  that parameters  to be
free.  The F-statistic values are 1.59, 1.51 and 1.08, respectively, for 120
additional parameters.   These results indicate that  the inclusion of
$\Gamma$ and $F_{K\alpha}$  as free parameters in the  orbital fits is
justified, but that  there is no justification for  including $R$ as a
free  parameter.  
On the  basis of  this test and the  simulations of
\S~\ref{simulations}, allowing $R$ to vary was not justified, so it 
was kept fixed at its time-averaged value in
the  following  spectral fits.   
This  situation  might be  physically
realized if the reflecting material is very close to the X-ray source.
The  inclusion  of  line energy  as  a  free  parameter was  also  not
justified.  None of the spectral models used in this analysis included
a constant,  narrow component to the  iron line as  neither the \asca\
observations nor  the \chandra\  observations suggest the  presence of
such a feature (Lee \et 2000b).

Lee  \et   (1999)  suggest  that  the  elemental   abundances  of  the
reprocessor in  \mcg\ differ from  solar.  To check that  the spectral
variability  results  presented here  are  not  an  artifact of  using
incorrect element abundances, the  spectral fitting was repeated using
the abundances of  Lee \et (1999).  Specifically, iron  was assumed to
be  over-abundant   by  a  factor   2  while  lighter   elements  were
under-abundant  by the  same factor.   Using this  model  the absolute
values  of the  fit parameters  changed  slightly but  the pattern  of
variability described below was essentially unchanged.

\subsection{ Error Estimation } 
\label{errors}

Fitting this  simple model to  all 120 orbital spectra  gave unusually
low values of the fit-statistic, with $\chi_{\nu}^{2} < 1.0$ in 114 of
the fits.  This is  most likely a result of the error  bars on the net
(source minus background)  spectrum being overestimated.  As discussed
by Nandra \et  (2000), the standard software calculates  errors on the
net spectrum by  combining errors on the total  spectrum with those on
the  model  background spectrum.  The  errors  on  the background  are
calculated  using Poisson  statistics on  the background  model, which
will tend  to overestimate the  true uncertainty on the  background in
the case  of short exposures, since the model is derived  from a large
amount of blank-sky data. For the purpose of assessing the variability
of  the  measured parameters  of  \mcg,  all  errors produced  by  the
standard software have been ignored and new ($1 \sigma$) errors were calculated
based on  the properties of the  derived light curves.   The errors on
the source count  rate were derived from the error on  the mean of the
$\sim 150$ individual 16s points within each orbital bin.

For the  spectral fit parameters,  where 16s data were  not available,
the  errors have  been calculated  based  on the  properties of  the
orbital light  curves.  Specifically, the error, $\sigma$,  on a given
parameter was calculated from the point-to-point variance of the light
curve for that parameter using the following equation:

\[ \sigma^{2} = \frac{1}{2} \frac{1}{N-1} \sum_{i=1}^{N-1} (x_{i} - 
x_{i+1})^{2} \]

\noindent where $x_i$ are the $N$ points in the light curve.

Simulations of constant light  curves with known errors confirmed that
the above equation  gives an accurate representation of  the errors in
this  situation (see  also \S~\ref{simulations}).   These are  in fact
conservative error estimates: in the absence of intrinsic variability,
the point-to-point scatter in a light curve will be due only to 
measurement errors.   If the source  has intrinsic variations  on such
short time  scales, the true error  will be smaller  than this because
part of the scatter is due to those intrinsic variations.  However,
even these conservative error estimates cannot account for systematic
errors.
Among these are the unlikely but conceivable possibilities that  rapid  
changes  in the  spectral response of  the PCA, or in  the energy and  
depth of an iron  K edge (from e.g., the warm  absorber) could mimic  
the observed line  variations.   

The  error on  the  power-law slope  derived  from the  point-to-point
scatter was $\sigma_{\Gamma} = 0.059$,  which compares to a mean error
of 0.028 from the standard software.  For the iron line flux the error
derived  was  $\sigma_{Fe}  =  3.4 \times  10^{-5}$  photon  cm$^{-2}$
s$^{-1}$,  compared  to  $4.3   \times  10^{-5}$  from  the  standard
software.

\subsection{ Spectral Variability Results } 
\label{correlations}

\begin{figure}\vspace{15 cm}
\includegraphics{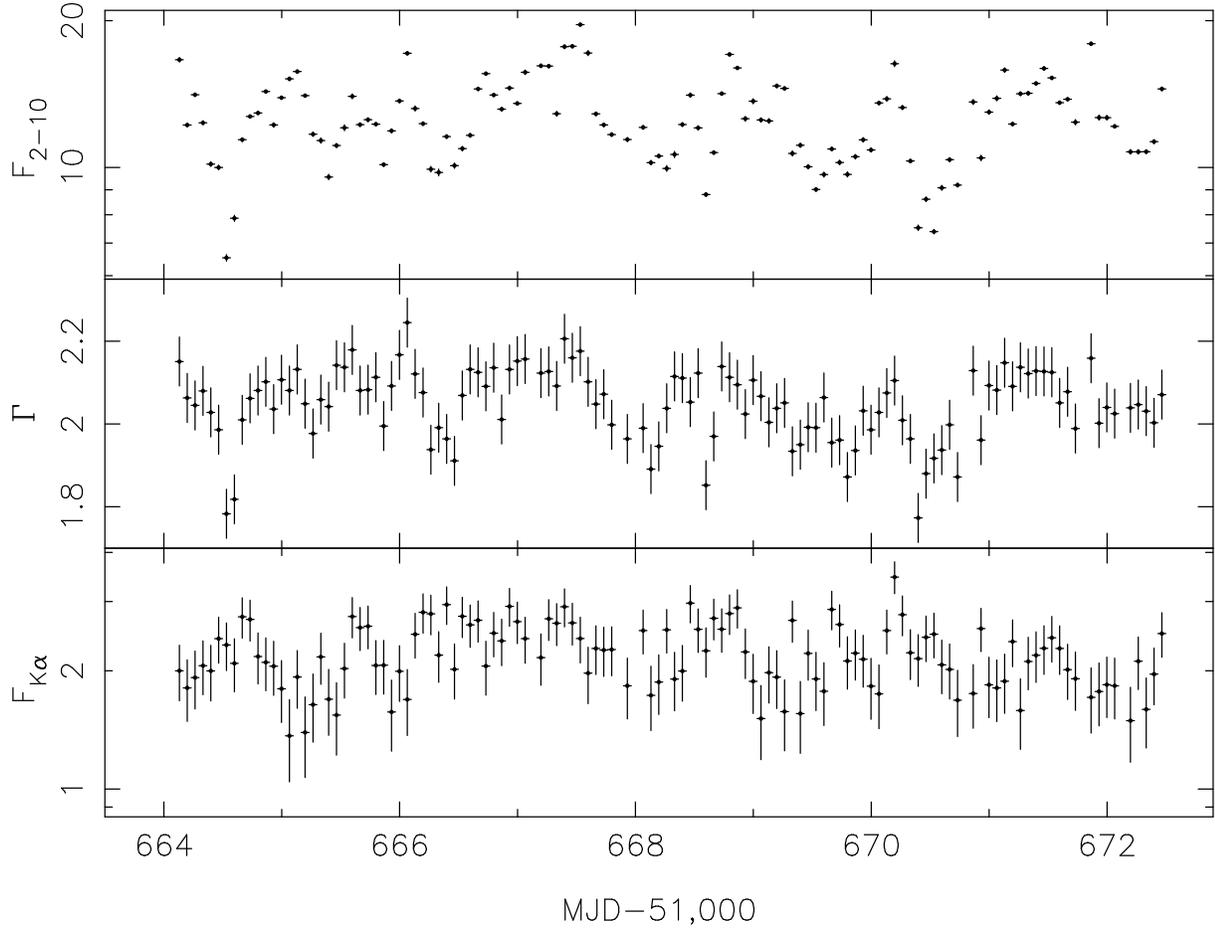}
\caption{ 
Light curves of the fit parameters derived by fitting the 120 orbital
spectra with a model comprising a power-law, a Gaussian emission line
and a reflection component (with fixed $R$). The spectral model is
described in \S~\ref{fits} and the $1 \sigma$ error bars are derived in
\S~\ref{errors}. The top panel shows the 2--10~keV light curve of
\mcg\ (in ct s$^{-1}$) binned to match the orbit of \xte\ (orbital 
period $\approx
96$~min). The other panels show the power-law slope $\Gamma$ and the
iron line flux ($10^{-4}$ photons cm$^{-2}$ s$^{-1}$). Note
the log scale on the $y$--axes.}
\label{lc}
\end{figure}

The light curves derived by  fitting all 120 orbital spectra are shown
in  Figure~\ref{lc}, with  error bars  calculated as  in  the previous
section.   The  top  panel  shows  strong,  rapid  variations  in  the
2--10~keV   count  rate,   $F_{2-10}$,   which  was   assumed  to   be
representative  of the  continuum flux  of the  source (see  Nowak \&
Chiang 2000 and  Lee \et 2000a for detailed  analyses of this light
curve).   The   power-law  slope  appears   significantly  variable;  a
$\chi^{2}$--test against a constant hypothesis gave $\chi^{2}$=261.2 /
119 {\it dof}, corresponding to a rejection of the constant hypothesis
at  99.99~\% confidence.  The  iron line  flux also  shows significant
variability,  with $\chi^{2}$=196.1  / 119  {\it dof}  (significant at
99.9~\%).  Again,  it is  noted  that  these  $\chi^{2}$ values  were
calculated using the conservative errors discussed earlier.

\begin{figure}\vspace{10 cm}
\includegraphics{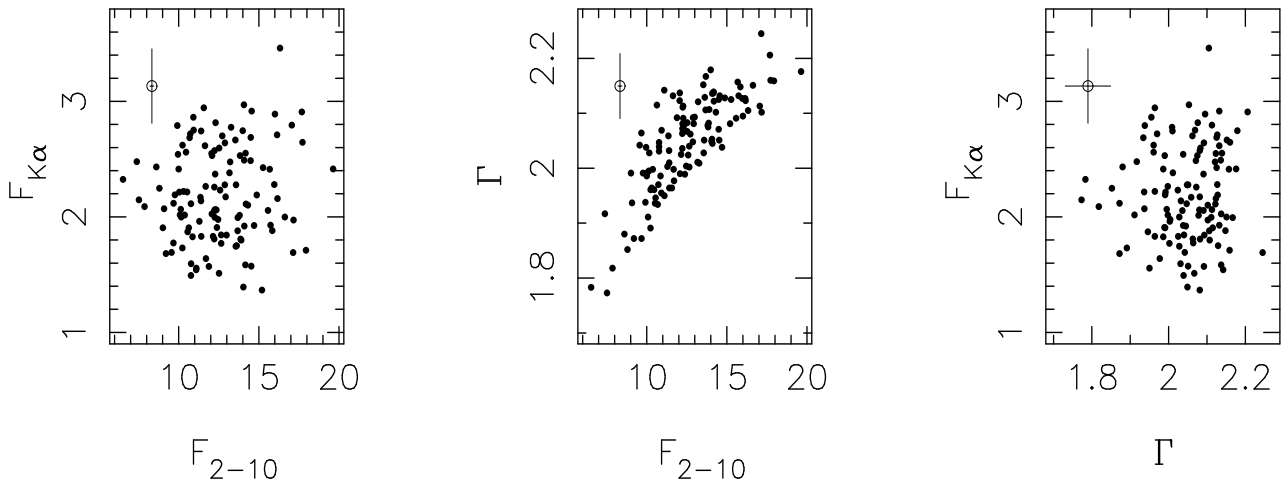}
\caption{ 
Zero-lag correlation diagrams for the data shown in figure~\ref{lc}.
The size of the error bars, calculated in \S~\ref{errors}, are indicated in the top left corner of
each panel. The only strong correlation observed is between $\Gamma$ and
$F_{2-10}$. 
}
\label{correl2}
\end{figure}

Figure~\ref{correl2} shows the zero-lag correlation diagrams for these
light curves.  Two non-parametric statistics were employed to test for
correlations, namely  the Spearman rank-order  correlation coefficient
$r_{s}$  and the  Kendall $\tau_{K}$--statistic  (see e.g.,  Press \et
1992).  Only one correlation  was significant, that between $F_{2-10}$
and  $\Gamma$, with $r_{s}=0.778$  and $\tau_{K}=0.587$,  which differ
from  zero  with  significance  levels $<10^{-20}$.   The  correlation
coefficients for  $F_{2-10}$ against $F_{K\alpha}$  were $r_{s}=0.076$
and $\tau_{K}=0.043$ and for  $\Gamma$ against $F_{K\alpha}$ they were
$r_{s}=0.056$ and $\tau_{K}=0.037$, consistent with no correlation 
below the 40\% level.

This tendency  for the  X-ray spectrum to  steepen as the  source gets
brighter has been observed in  many Seyfert 1s (e.g., Nandra \et 1991;
Ptak \et 1994; Nandra \et 1997a; Markowitz \& Edelson 2000).  However,
it has not been clear whether  this is due to intrinsic changes in the
slope of the  underlying power-law or the superposition  of a constant
reflection  component on  a power-law  of constant  slope  but varying
normalization.   The inclusion  of (fixed $R$) reflection in  the above  spectral
fitting  suggests that  the  $\Gamma$-flux correlation  is genuine  in
\mcg.  As a check the  above fits were  repeated using a reflection  component of
fixed absolute  flux (i.e., slope  and normalization of  the power-law
incident on the reflector held  constant), as might be expected if the
reflecting  material  is  very   distant  and  responds  only  to  the
time-averaged continuum.   The fits again show  a strong $\Gamma$-flux
correlation.  

The correlation  diagrams of  Figure~\ref{correl2} show that  the iron
line flux  is not correlated with  the source flux  nor with $\Gamma$.
The rapid  variations in  the line flux  do however suggests  that the
majority of the  line flux does indeed originate  in a compact region.
Indeed,  while  the iron  line  is not  as  strongly  variable as  the
continuum,  it does  appear variable  on time  scales  $\sim 0.5-2$~d,
which is the shortest time for a factor of two change (e.g. day 665). 
In order to search for  possible time-delays between the continuum and
line,  temporal cross-correlation functions  were computed  using both
the interpolated  correlation function  (ICF; White \&  Peterson 1994)
and the discrete correlation function (DCF; Edelson \& Krolik 1988).

\begin{figure}\vspace{12 cm}
\includegraphics{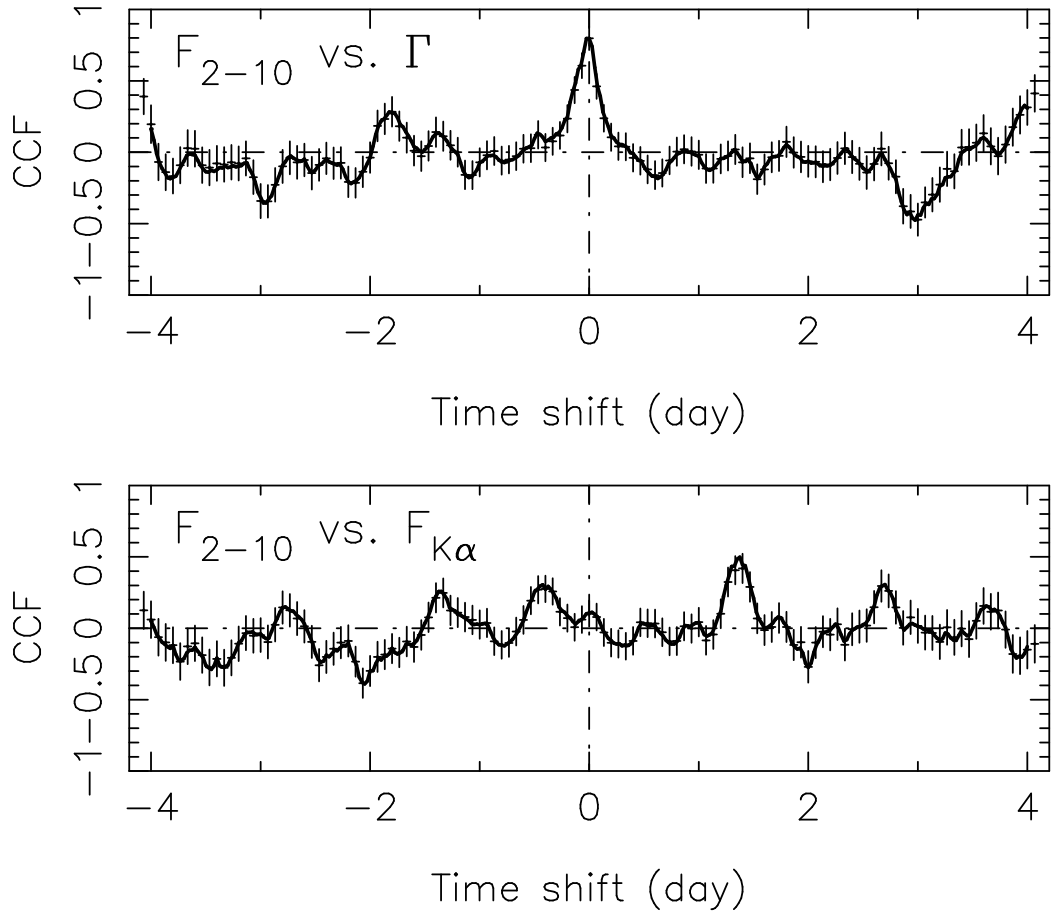}
\caption{ 
Cross-correlation functions for $F_{2-10}$ against $\Gamma$ (top
panel) and against $F_{K\alpha}$ (bottom panel).
The solid line refers to the interpolated correlation function while 
the error bars refer to the discrete correlation function.
Only $F_{2-10}$ against $\Gamma$ shows a significant correlation.
}
\label{ccf}
\end{figure}

Figure~\ref{ccf}  shows  the  cross-correlation functions.   In  these
plots a  positive time shift  corresponds to the first  quantity (i.e.
$F_{2-10}$) leading  the second.   Both the ICF  and DCF  for 
$F_{2-10}$ against $\Gamma$  reach maximum correlation coefficients of
$r_{max}=0.80$ at zero-lag.   The correlation functions for $F_{2-10}$
against $F_{K\alpha}$ reach peaks at $\tau = +1.4$~day (continuum flux
leading  line),  although  in   both  cases  the  maximum  correlation
coefficient is  small ($r_{max}=0.50$  and 0.42 for  the ICF  and DCF,
respectively).

The apparent  lack of a  strong correlation between the  continuum and
line fluxes  is difficult  to explain in  the context of  the standard
model, as discussed in \S~\ref{discussion}.

\section{ Simulated Spectra } 
\label{simulations}

In order to assess the robustness of this spectral analysis, simulated
data were produced  to resemble as closely as  possible the real data,
and  then analyzed in  an identical  fashion.  Each  simulated dataset
consisted of 120 spectra  produced using spectral models together with
response matrices,  exposure times  and background spectra  taken from
the  real data.  Poisson  noise was  added to  the total  (source plus
background) spectra  and the entire simulated dataset  was then fitted
in exactly the  same fashion as the real data.   As the intention here
was to produce simulated data which matched the properties of the real
data as closely as possible, the original background files, which were
derived from a  model, not observed photons, were  used in the fitting
without  Poisson noise added.   This meant  that net  spectra produced
from the  simulated data  also had systematically  overestimated error
bars (see  \S~\ref{errors}).  For  each test 10  complete sets  of 120
spectra were produced.

\subsection{ Iron Line Simulations } 
\label{line_sim}

The spectral  models used to  produce the simulated data  were created
with parameters derived from fits  to the real data.  For example, the
first set of simulated data contains a variable continuum but constant
iron line.  The  spectral models used to produce  these simulated data
were obtained by fitting the  real data, then setting $F_{K\alpha}$ to
its time-averaged  value before the simulated  spectra were generated.
Thus, the simulated data should have had spectra with continuum slopes
and normalizations  that varied  like the real  data, but  line fluxes
that remained constant.

\begin{figure}\vspace{10 cm}
\includegraphics{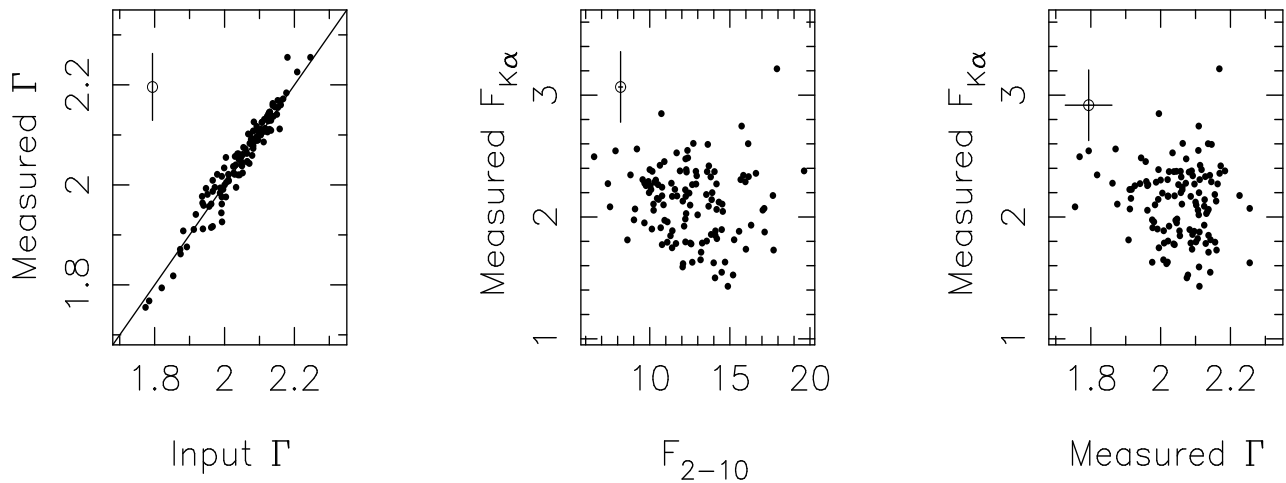}
\caption{ 
Correlation diagrams for a simulated dataset. 
The values of $\Gamma$ and $F_{2-10}$ were taken from
the real data but $F_{K\alpha}$ was constant. 
Again the error bars are calculated as in \S~\ref{errors}.
There are no spurious correlations between the measured parameters and
$F_{K\alpha}$ is consistent with constant (the scatter is much smaller
than in Figure~\ref{correl2}).
}
\label{simul2}
\end{figure}

These  constant-line  simulations  were  used  to  justify  the  error
estimation of  \S~\ref{errors}.  With  the errors calculated  from the
point-to-point scatter,  each simulation gave  $\chi_{\nu}^{2} \approx
1.0$  against a  constant  hypothesis (from  10  simulations the  mean
$\chi_{\nu}^{2}$  was 1.01 with  a standard  deviation of  0.09), indicating
that  this  method  does  indeed   provide  a  good  estimate  of  the
measurement error.  The difference in $\chi^{2}$ between the real data
and  the   constant-line  simulations  suggested   that  the  measured
variability in the line is highly significant ($\chi_{\nu}^{2} = 1.64$
for  the  real   data).   Figure~\ref{simul2}  shows  the  correlation
diagrams from one of the constant-line simulations.

With the errors calculated using  the standard software there is still
a significant difference  in $\chi^{2}$ between the real  data and the
constant line  simulations.  Performing the  $\chi^{2}$--test as above
but using the  standard errors gave a $\chi_{\nu}^{2}  = 1.05$ for the
real  data and  $\chi_{\nu}^{2}  \approx 0.65$  for the  constant-line
simulations.   As the  light curve  for  a constant  line should  have
$\chi_{\nu}^{2} =  1 $ if  the errors are  correct, the fact  that the
simulations gave  $\chi_{\nu}^{2} \approx 0.65 $  again reinforces the
view that the standard  errors are overestimates, and the recalculated
errors are more accurate.

\begin{figure}\vspace{14 cm}
\includegraphics{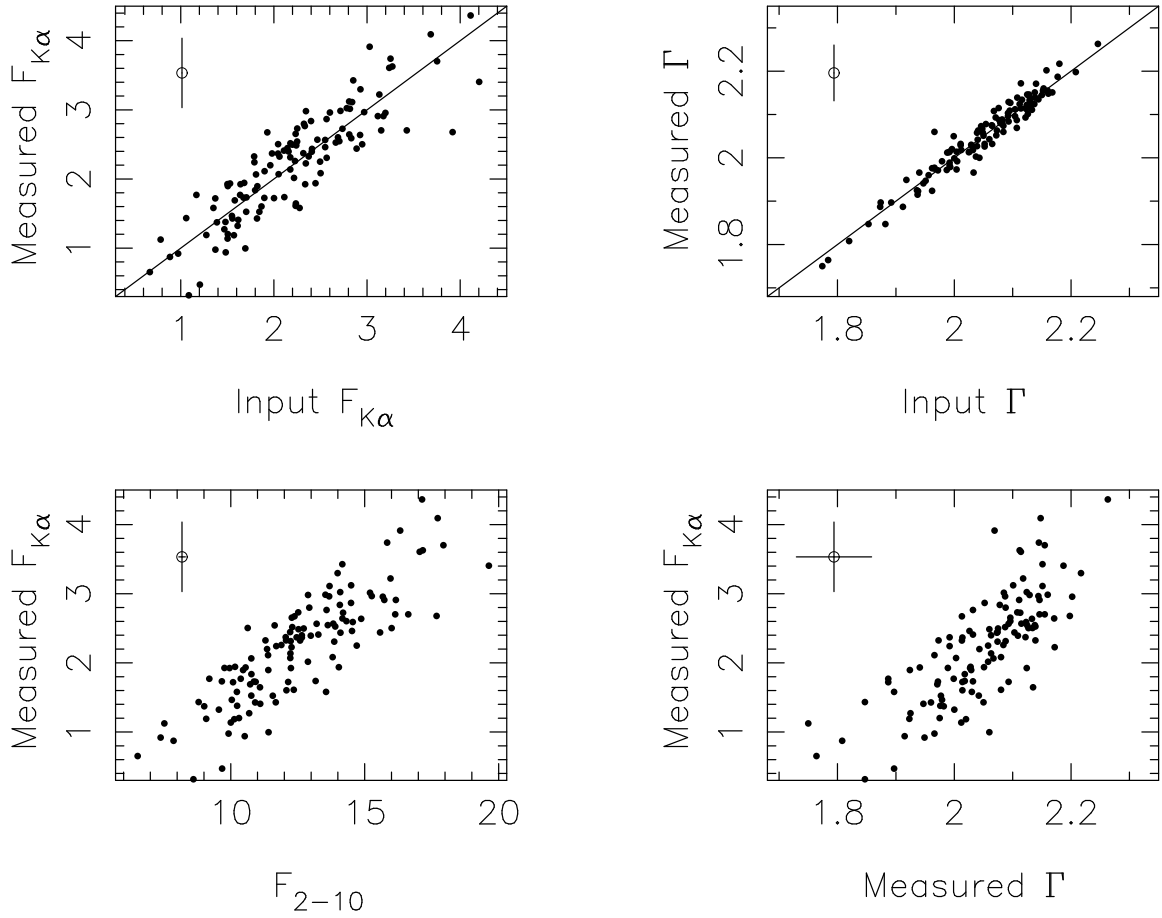}
\caption{ 
Correlation diagrams for a simulated dataset. 
The values of $\Gamma$ and $F_{2-10}$ were taken from
the real data and $F_{K\alpha}$ is proportional to $F_{2-10}$. 
The correlation between $F_{K\alpha}$ and $F_{2-10}$ and is clearly well recovered
(and as a result, a secondary correlation between $F_{K\alpha}$ and
$\Gamma$ is apparent).  
}
\label{simul3}
\end{figure}

Simulations have  also been generated  in which the line  flux follows
the  continuum flux  with zero  lag (i.e.,  $F_{K\alpha}$ is  a linear
function  of  power-law  normalization).   In  these  simulations  the
equivalent width  of the line is approximately  constant.  Again, fits
to  these simulated  data recovered  the input  values  accurately and
this time showed the expected strong  correlation between  line flux  
and continuum flux (see Figure~\ref{simul3}). 

These  simulations suggest that  the two  simplest hypotheses  for the
behavior  of the  iron  line, namely  a  constant flux  or a  constant
equivalent width, can be rejected in the case of \mcg.

\subsection{ Continuum Simulations } 
\label{cont_sim}

\begin{figure}\vspace{13.5 cm}
\includegraphics{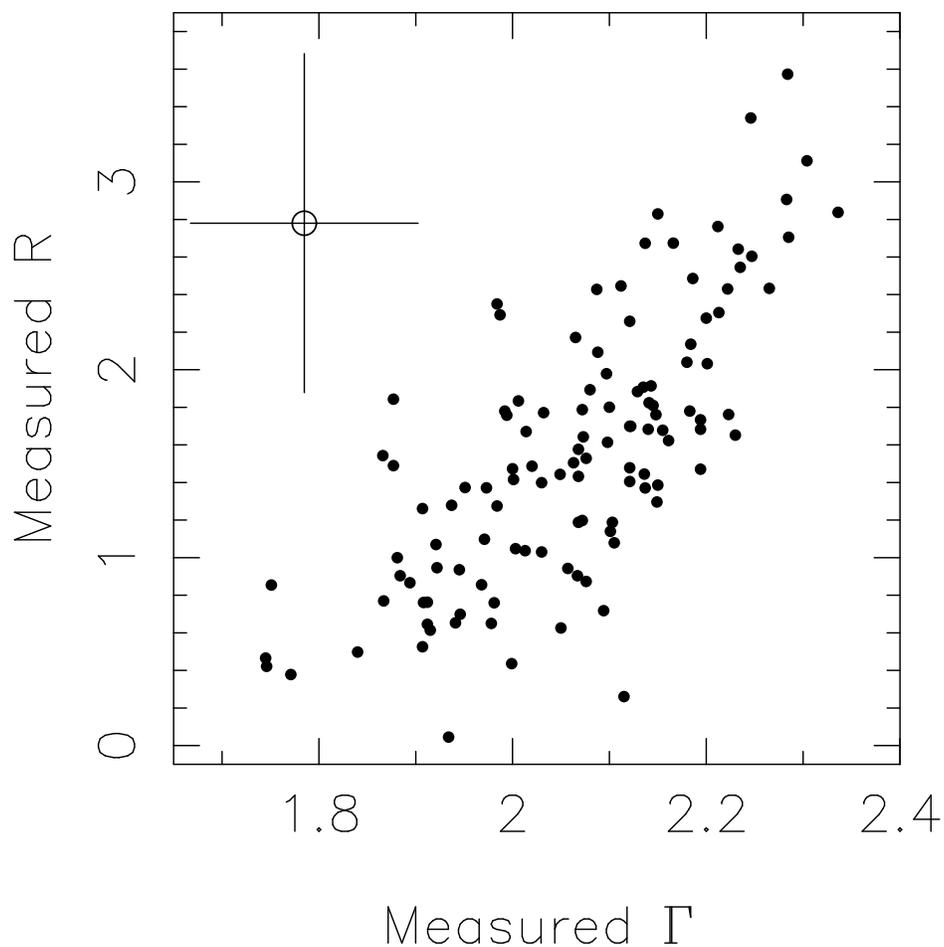}
\caption{ 
Correlation between the values of $\Gamma$ and $R$ measured
from simulated data with constant $R=1.42$.
As $R$ is fixed, there should be no correlation between $\Gamma$ and 
$R$.
The errors have been determined using
the conservative approach outlined in \S~\ref{errors}.
The fact that there is a clear correlation indicates a bias in the 
simultaneous determination of $R$ and
$\Gamma$, and thus any observed correlation between $\Gamma$ and $R$ 
must be considered with caution. 
}
\label{simul4}
\end{figure}

These simulations also reveal a bias in the simultaneous determination
of $\Gamma$ and $R$  from these data.  Figure~\ref{simul4} shows the
results of fitting a simulated  dataset in which the model (input) $R$
was held constant.  
In  order to test how accurately  $\Gamma$ and $R$
can be simultaneously constrained,  $R$ was left as a
free parameter When fitting the  simulated data in this test.
There  is a clear tendency for the
value of  $R$ measured  from the fit  to increase with  $\Gamma$.  The
correlation coefficients are  $r_{s}=0.739$ and $\tau_{K}=0.554$, both
indicating  that  $\Gamma$  and  $R$  are correlated  with  very  high
significance.  As $R$ is known to be constant in these simulated data,
this can  only be  an artifact of  the fitting procedure.   Chiang \et
(2000) and Nandra \et (2000)  have previously discussed this effect in
the context of their respective spectral variability analyses.  On the
basis of these simulations it is concluded here that this effect is an
artifact of the interdependency of the two model components, power-law
and reflection, and that it  is not possible to simultaneously measure
both $\Gamma$ and $R$ from the  real data in an unbiased fashion.  The
implications  of  this  bias  for  previous studies  is  discussed  in
\S~\ref{discussion}.

A secondary point is that such a correlation introduced by the fitting
procedure  is  not the  source  of  the  observed correlation  between
$\Gamma$ and  $F_{2-10}$, which persists independent  of the inclusion
of fixed (absolute or relative) reflection in the fit.

\section{ Discussion } 
\label{discussion}

This paper  presents time-resolved X-ray spectroscopy of  \mcg\ with a
resolution of 96~min.  On these timescales there is evidence that the
iron line  flux is  variable but does not
appear to be simply related to the observed continuum flux.  Also, the
slope of  the underlying power-law continuum  varies significantly and
appears to  be correlated with the  source flux in the  sense that the
power-law  steepens  as the  source  luminosity increases.   Extensive
simulations have been  used to assess the robustness  of these results
and search  for possible sources  of error in the  line determination;
none  were found.   That said,  it  is not  possible to  rule out  the
influence of  subtle systematic effects with  arbitrary confidence, as
discussed earlier. 
While these  seem  unlikely,  confirmation of  the
observed iron line  variability must come from missions  such as \xmm\
which offer high  throughput and spectral resolution far  in excess of
\xte.   However, for  the  purpose of  the  following discussion,  the
measured line variations are assumed to be intrinsic to the source.

\subsection{ Reconciliation with Previous Analyses }

The temporal properties of the iron line in \mcg\ have previously been
addressed by  Iwasawa \et  (1996, 1999), Lee  \et (2000a)  and Reynolds
(2000).   With the  line properties  measured  as a  function of  time
throughout an observation  (e.g. \S~3.2 of Iwasawa \et  1996) the line
flux appeared variable on timescales $\sim 4 \times 10^{4}$.
However,  when  the 
observations were  broken into intervals  of high and low  fluxes they
did not show the expected
correlation  between the  line  flux and  continuum, and  even
variability   of  the   line  was   difficult  to   detect   (Lee  \et
2000a). Reynolds  (2000) used an interpolation procedure  to search for
lags or  leads between a continuum  band (2--4~keV) and  the line band
(5--7~keV) but found none.

The  fully time-resolved  spectral  analysis presented  in this  paper
suggests that  the line flux is  indeed variable on  short timescales
($\sim  1$~d)  but is  not  correlated  with  the continuum  over  the
duration of the observation, thus  accounting for the apparent lack of
line  variability seen  in  the flux-correlated  analyses.  A  further
point  is that  the lack  of correlation  between line  and continuum,
combined  with  smaller amplitude  variations  in  the  line than  the
continuum,  is  responsible  for  the  anti-correlation  between  line
equivalent width and continuum luminosity observed by Lee \et (2000a).

\subsection{ Rapid Iron Line Variability }

This  lack  of a  correlation  between  the  line flux  and  continuum
suggests that  the reprocessing  is more complex  than that  needed to
produce a  strong reverberation  signature (Blandford \&  McKee 1982).
Some obvious possibilities that could  affect the response of the line
include  the  non-negligible  size  of the  X-ray  emitting  region(s)
compared to the  inner disk, the presence of  a strongly ionized layer
on  the surface of  the disk  (Ross, Fabian  \& Young  1999; Nayakshin
2000), the orbital motion of both the X-ray emitting region and the
disk material  (Ruszkowski 1999), and  the geometry of the  inner disk
(e.g. Blackman 1999).

Perhaps the most  promising of these is the effect  of an ionized skin
on the  properties of the iron  K$_{\alpha}$ line.  It  is likely that
the  bulk  of  the  line  flux  in \mcg\  originates  in  the  deeper,
relatively neutral layers  of the disk since the  line appear neutral.
Scattering in an overlying, ionized skin can be effective at degrading
the strength  of the neutral iron  line, as discussed  by Nayakshin \&
Kallman  (2000) and  Nayakshin  (2000).  The  temperature and  optical
depth of  the ionized  skin, which are  functions of the  geometry and
spectrum  of  the X-ray  source,  thus have  a  strong  effect on  the
observed strength of  the line.  It is interesting  to note this means
it is in principle possible for  the line flux to change without there
being any  change in the continuum  over the \xte\  bandpass.  This is
because  the temperature  of the  ionized skin  is a  function  of the
energy  at  which the  X-ray  spectrum  rolls  over (see  \S~6.3.2  of
Nayakshin  \& Kallman  2000)  and the  strength  of ultraviolet  bump,
neither of which  are observed by \xte\ but could  be (and most likely
are) variable.  Alternatively the physical conditions of the accretion
disk   skin   could  be   governed   by   some   process  other   than
photoionization. One  strong candidate is mechanical  heating from the
magnetohydrodynamic turbulence  intrinsic to  the disk, which may
also be responsible for the  formation of the disk corona (e.g. Miller
\& Stone 2000).

Based on the X-ray variability  of \mcg, Iwasawa \et (1996) and Fabian
(1997) suggested that the X-ray source at any given time consists of a
few discrete flaring regions each illuminating a relatively small area
of the disk.   This `patchy' geometry makes modelling  the response of
the line to the continuum difficult, especially if the flaring regions
are close to the disk surface (Nayakshin 2000).  Directly following an
X-ray  flaring event  the  accretion  disk surface,  at  least in  the
immediate vicinity  of the flare, is  likely to be  out of hydrostatic
equilibrium,  and  the  timescale  for  the  material  to  return  to
equilibrium is long compared to the light-crossing time (Nayakshin \et
in prep.). During periods when  the disk surface is out of equilibrium
the line emission would be poorly  correlated with the driving continuum due
to  ionization  instabilities.   The  line  emission  may therefore only
correlate with  the continuum emission on timescales  much longer than
the expected  light-crossing time, which will average  out the effects
of  orbital modulation and  instabilities. For  example, if  the source
were to  enter a prolonged  low state (as  recently seen in  NGC 4051;
Uttley  \et   1999)  then  the   iron  line  flux   should  eventually
respond. Thus it is important  to measure spectral variability on long
as well  as short time scales  in order to understand  the response of
the  iron $K_{\alpha}$  line  to the  continuum.   An alternative,  if
rather speculative explanation for the unusual line variations,, is
that one of  the basic assumptions 
about  the production  of  the broad  iron  line by  reflection off  a
geometrically thin, optically thick accretion disk is flawed.

\subsection{ Continuum Slope Variability }

The strong correlation between  X-ray flux and underlying spectral slope
found in  \mcg\  can  be  disentangled from  the effects  of
reflection.   Similar  effects  have been
observed in NGC 5548 (Chiang \et 2000), NGC 7469 (Nandra \et 2000) and
IC 4329a (Done,  Madejski \& \.{Z}ycki 2000), although  in the case of
NGC  7469,  at least  on  timescales  of  days,  the power-law  slope
appeared better  correlated with  the ultraviolet luminosity  than the
X-ray luminosity.

These results  imply that some  property of the X-ray  emitting corona
(e.g.  electron temperature, optical depth, size) is changing with the
X-ray flux, although at this stage it is not clear which one it is.  Various
authors  have  investigated  the  spectral variability  properties  of
accretion  disk coronae for  both AGN  and Galactic  accreting sources
(e.g. Haardt,  Maraschi \& Ghisellini  1997; Poutanen \&  Fabian 1999;
Malzac \& Jourdain 2000).  In these models the simplest way to produce
a slope-flux correlation  similar to that observed in  \mcg\ is if the
X-ray variability is  driven by changes in the  seed photon population
(e.g.   \S~5  of  Malzac  \&  Jourdain  2000).   This  possibility  is
consistent  with the tentative  claim that  the EUV  flux in  NGC 5548
leads the hard X-ray flux by $\sim 10 - 30$~ks (Chiang \et 2000).

\subsection{ Compton Reflection and Other Spectral Components }

Another use of spectral variability  to study AGN physics is to search
for  a  relation  between  the  strength  of  the  Compton  reflection
component  and   other  properties.   The  strength   of  the  Compton
reflection  component  could   in  principle  provide  an  independent
diagnostic of  the geometry/albedo of  the disk.  A recent  example is
the  result of  Zdziarski, Lubi\'{n}ski  \& Smith  (1999), who  find a
correlation between $\Gamma$ and $R$ measured in a  sample  of  Seyfert  galaxies.   If
confirmed, such a correlation  would suggest that there is significant
feedback between the X-ray source and the reprocessing medium.

Unfortunately,  the \ginga\  and \xte\  data gathered  to  date cannot
provide good  constraints on $R$ on  short timescales.   In fact, the
simulations of \S~\ref{cont_sim} show that  there is a serious bias in
the simultaneous  determination of  $\Gamma$ and $R$  that leads  to a
strong,  false correlation  between these  two parameters.   A similar
effect  was discussed by  Chiang \et  (2000), Nandra  \et (2000)  and Matt
(2000).   The spurious  correlation found  in the  simulations  in the
current  paper  (see  \S~\ref{cont_sim})  spans almost the  entire  range  of
parameter space  probed by  Zdziarski \et (1999)  in their study  of a
sample of Seyfert galaxies, and indicates that the question of whether
$\Gamma$ and $R$ are intrinsically  correlated, as suggested to be the
case  by those  authors, remains  open.  
A corollary of this is that other
correlations involving $R$ may be biased by this effect. For example,
the anti-correlation between line equivalent width and $R$ noted by
Chiang \et (2000) and Lee \et (2000a) may be due to a combination of
an intrinsic $\Gamma$-flux correlation, a relatively constant iron
line flux and a spurious correlation between $\Gamma$ and $R$. In
this case the line equivalent width would appear to anti-correlate
with $\Gamma$ and hence with $R$.
These  results  highlight the
difficulties in simultaneously  fitting two broad continuum components
from data of limited band-pass.
A final point is that the line measurements shown in Figure~2 remain
largely unchanged if $R$ is left free in the spectral fitting. The
main results of this paper are therefore not strongly affected by the
presence of a genuine correlation between $R$ and $\Gamma$ (which
would be accounted for in the free-$R$ fits).

\section{Conclusions} 
\label{conclusions}

This  paper  presents  a  new  method  of  analyzing  Seyfert~1  X-ray
variability  data, designed  to measure  rapid spectral  variations by
fitting data from individual orbits.  This was applied to a long (8~d)
\xte\ observation  of the bright,  strongly variable Seyfert  1 galaxy
\mcg,  and the  conclusions  were tested  with extensive  simulations.
This provided  the clearest evidence to  date that the  iron line flux
($F_{K\alpha}$) of  a Seyfert~1 varies on  relatively short ($\ls$1~d)
timescales.   However, the variations were {\it  not} correlated with
or simply  related to the continuum  variations.  As the  line is very
broad, indicating that  the bulk of the flux is  produced close to the
black hole,  the lack of correlation contradicts  the simplest picture
in which the line variations  are driven by variations in a point-like
continuum  source, modulated by  light-travel effects.   Although many
modifications to  this standard  picture are possible,  one reasonable
solution is  that the  line  variations  are
dominated   by   some   other   process,  for   instance,   ionization
instabilities, in which case the dominant timescale would be that for
the  system to  return to equilibrium.   Another  is that
there is something wrong with the simple disk-corona picture described
above.   In  any event  this  result  indicates  that the  physics  of
reprocessing  in the central  engine is  more complex  than previously
thought.

The slope  of the underlying  X-ray power-law ($\Gamma$) shows strong
variability  and is  tightly correlated  with the  source flux  in the
sense that the  power-law steepens as the source  gets brighter.  

This behaviour  indicates that the  properties of  the  corona are
rapidly variable and, if Compton-cooling is the dominant process in
determining the X-ray spectral slope,  may be driven  by variations in
the  unobserved EUV  band. The limited band-pass  of these data  and intrinsic
bias  in the  simultaneous measurement  of $\Gamma$  and  $R$ appear to conspire
to produce a spurious  correlation between these two quantities.  This
strongly supports the conclusions  of previous works (e.g., Nandra \et
2000) that  indicates that  the correlation  claimed in  Zdziarski \et
(1999) may be an artifact of the analysis.

The unexpected result  concerning the iron line variability  hint at much
more complex  behavior than  expected in Seyfert  1s.  \xmm,  with its
unprecedented combination of  high throughput and spectral resolution,
is ideally  suited to study  rapid iron line variability.   We eagerly
await the  results of  the long observations  of this and other 
Seyfert 1s  that will soon be made by \xmm.

\acknowledgments 
It is  a pleasure to thank Tess  Jaffe and the members of the \xte\
GOF for assistance
with   the  data  reduction,   Alex  Markowitz   for  help   with  the
cross-correlation  functions  and  Keith  Arnaud for  help  with  {\tt
XSPEC}.  We are also grateful  to Sergei Nayakshin, Bob Warwick and
Chris Reynolds for
valuable discussions.   This research made  use of data  obtained from
the   High  Energy  Astrophysics   Science  Archive   Research  Center
(HEASARC),  provided  by  NASA's  Goddard  Space  Flight  Center.   SV
acknowledges support from PPARC and RE acknowledges support from NASA
grants NAG~5-7317 and NAG~5-9023.

\pagebreak

\pagebreak

\end{document}